\documentclass[pra,twocolumn]{revtex4}
\usepackage{amsmath,amssymb,bm}
\usepackage{color}

\usepackage{graphicx}
\usepackage{cases}

\begin{document}\title{Finite Temperature Properties of Three-Component Fermion Systems in Optical Lattice}
\author{Hiromasa Yanatori}
\affiliation{Department of Physics, Tokyo Institute of Technology, 
Meguro-ku, Tokyo 152-8551, Japan}
\author{Akihisa Koga}
\affiliation{Department of Physics, Tokyo Institute of Technology, 
Meguro-ku, Tokyo 152-8551, Japan}
\date{\today}

\begin{abstract}
We investigate finite temperature properties 
in the half-filled three-component (colors) fermion systems. 
It is clarified that a color density-wave (CDW) state is more stable than
a color-selective "antiferromagnetic" (CSAF) state 
against thermal fluctuations.
The reentrant behavior in the phase boundary for the CSAF state is found.
We also address the maximum critical temperature
of the translational symmetry breaking states 
in the multicomponent fermionic systems.
\end{abstract}


\maketitle
\section{Introduction}
Ultracold fermions have attracted much 
interest~\cite{ultracold1,ultracold2,ultracold3}.
One of the interesting topics is the phase transition to 
the symmetry breaking states such as the superfluid and 
magnetically ordered states.
The former has been observed in the optical lattice~\cite{Chin},
and the BCS-BEC crossover has also been 
discussed~\cite{crossover1,crossover2,crossover3,crossover4,crossover5}.
On the other hand, the latter translational symmetry breaking state 
should be hard to be realized since intersite correlations are
extremely small in optical lattice systems.
Recently, it has been reported that two component fermions 
reach a very low temperature 
close to the Neel temperature $\sim 1.4 T_N$~\cite{1.4T_N},
which should accelerate further theoretical and experimental investigations
on the observations of the translational symmetry breaking state.

Multicomponent fermion systems such as 
Li~\cite{Li1,Li2}, Yb~\cite{Yb1,Yb2} and Sr~\cite{Sr1,Sr2}, 
should be the possible candidates to observe 
the translational symmetry breaking states.
Miyatake et al. have theoretically studied ground state properties 
in the three component fermion systems with anisotropic interactions 
to clarify the existence of 
translational symmetry breaking states~\cite{Miyatake}. 
However, the stability of these ordered state against thermal fluctuations 
has not been discussed systematically~\cite{InabaReview}.
In particular, it is necessary to clarify whether or not these ordered states
can be realized at accessible temperatures.
In addition, it is desired to clarify the role of the multicomponents
in realizing the translational symmetry breaking state at finite temperatures.

Motivated by this, we consider the ultracold fermion systems with 
three components on the optical lattice.
Combining dynamical mean-field theory (DMFT)~\cite{DMFT1,DMFT2,DMFT3} with 
the non-crossing approximation (NCA)~\cite{Grewe,Kuramoto,eckstein},
we discuss finite temperature properties in the system.
We also study the translational symmetry breaking state
in the system with $N_c=2,3,\cdots, 6$, 
where $N_c$ is the number of components of fermions.
Then we demonstrate that the maximum critical temperature 
for the six-component system is about twice higher than
that for the two-component system.

The paper is organized as follows.
In \S \ref{2}, we introduce the model Hamiltonian for
the three component fermion systems on the optical lattice
and briefly summarize our theoretical approach.
In \S \ref{3}, we study how stable the competing ordered states are against
thermal fluctuations.
The transition temperatures for multicomponent systems are addressed
in \S \ref{4}.
A brief summary is given in the last section.

\section{Model and Method}\label{2}

We consider the three-component fermion systems with anisotropic interactions,
which should be described by the following Hubbard Hamiltonian,
\begin{eqnarray}
\hat{\cal{H}}=-t\sum_{\langle i,j \rangle ,\alpha}c^{\dagger}_{i\alpha}c_{j\alpha}+\frac{1}{2}\sum_{\alpha \neq \beta,i}U_{\alpha\beta}n_{i\alpha} n_{i\beta},
\end{eqnarray}
where $\langle i,j \rangle$ indicates the nearest neighbor sites
and $c^\dag_{i\alpha}$($c_{i\alpha}$) creates (annihilates) a fermion 
with color $\alpha$(=1,2,3) at site $i$ and 
$n_{i\alpha}=c^\dag_{i\alpha} c_{i\alpha}$.
$t$ is the hopping integral and $U_{\alpha\beta}(=U_{\beta\alpha})$ 
is the on-site interaction 
between colors $\alpha$ and $\beta$.
For simplicity, we set $U_{12}=U$ and $U_{23}=U_{31}=U'$.
Setting chemical potential 
$\mu_{\alpha}=\sum_{\beta\neq\alpha}U_{\alpha\beta}/2$,
we discuss the particle-hole symmetric systems.
In the case, the model Hamiltonian ${\cal H}(t, U, U')$ 
is transformed to ${\cal H}(t, U, -U')$ under the particle-hole 
transformations~\cite{Shiba} as $c_{i1}\rightarrow c_{i1}$,
$c_{i2}\rightarrow c_{i2}$, and $c_{i3}\rightarrow (-1)^ic_{i3}^\dag$.
Therefore, our discussions are restricted to the case $U'\ge 0$ without loss
of generality.

Low temperature properties for the Hubbard model in the infinite dimensions 
have been discussed so far. 
The existence of the Mott transitions and superfluid state
have been clarified~\cite{InabaMott,Inaba,Okanami}.
In addition to these, it is known that 
the translational symmetry breaking state is 
realizable in the bipartite lattice~\cite{Miyatake,InabaReview,Sotnikov1}.
In the particle-hole symmetric system, when $U<U'$, the repulsion between colors 1 and 2 is relatively smaller 
than the others, which leads to
the color density wave (CDW) state.
In the state, fermions with colors 1 and 2 are 
located at the sublattice $A$, and the others are located at sublattice $B$.
On the other hand, when $U'<U$, fermions with colors 1 and 2 
occupy alternatively 
in the different sublattices and fermions with colors 3 are itinerant.
Therefore, this state is regarded as the color-selective "antiferromagnetic" 
(CSAF) state~\cite{Miyatake}.
It is known that at zero temperature, 
the CDW state competes with the CSAF state and
the first-order quantum phase transition occurs on the SU(3) symmetric 
condition $U=U'$.


To discuss the stability of the above ordered states systematically,
we make use of DMFT~\cite{DMFT1,DMFT2,DMFT3}. In DMFT,
the many-body system is mapped to the single impurity model imposed on
the self-consistent condition. 
Here, we use the NCA method as an impurity solver,
where simple diagrams are involved~\cite{Grewe,Kuramoto,eckstein}.
Although the method may not be appropriate to discuss particle correlations
at very low temperatures, 
it has an advantage
in treating strong correlations in the fermionic systems 
with large degrees of freedom at finite temperatures~\cite{NCAwerner}.
Therefore, this method is complementary to 
the two-site approach~\cite{Twosite,SFA} 
used in the previous studies~\cite{Miyatake,InabaReview}, which
is appropriate in the weak coupling region\cite{SFA}. 

In this paper, we use a semicircular density of state (DOS) 
$\rho(\epsilon)=2\sqrt{(1-(\epsilon/D)^2}/(\pi D)$, 
where $D$ is the half-bandwidth.
When one considers the translational symmetry breaking state
in the bipartite lattice, 
the self-consistent equations~\cite{Chitra} are given by,
\begin{eqnarray}
{\cal G}_{\gamma\alpha}(i\omega_n)=i\omega_n+\mu-\left(\frac{D}{2}\right)^2G_{\bar{\gamma}\alpha}(i\omega_n),
\end{eqnarray}
where $G_{\gamma\alpha} (\cal{G}_{\gamma\alpha})$  
is the full (noninteracting) Green function with color $\alpha$ 
for the $\gamma(=A, B)$th sublattice.

To discuss how stable translational symmetry breaking states
are against thermal fluctuations,
we calculate 
the staggered parameter $M_\alpha$,
specific heat $C$ and entropy $S$, as
\begin{eqnarray}
M_\alpha&=&\frac{1}{N}\sum_i (-1)^i\langle n_{i\alpha}\rangle,\\
C&=&\frac{dE}{dT},\\
S&=&\int_0^T \frac{C}{T^\prime}\,dT^\prime,
\end{eqnarray}
where $N$ is the total number of sites and 
\begin{eqnarray}
E&=&E_K+E_U,\\
E_K&=&\left(\frac{D}{2}\right)^2\sum_\alpha\int_0^\beta\, d\tau \,G_{A\alpha}(\tau)G_{B\alpha}(-\tau),\\
E_U&=&\frac{1}{2N}\sum_{\alpha \neq \beta,i}
U_{\alpha\beta}\langle n_{i\alpha}n_{i\beta}\rangle.
\end{eqnarray}

Before we proceed our discussions, 
we specify some features for possible ordered states in the model.
In the CDW state, the staggered parameters have the relations 
$M_1=M_2\neq 0$ and $M_3 \neq 0$.
The CSAF state has itinerant fermions with color 3, and 
it is characterized by the relations $M_1=-M_2\neq 0$ and $M_3=0$.
The superfluid state~\cite{Inaba,Okanami} is also realizable in the model, 
and however the corresponding critical temperature 
obtained by DMFT with the NCA method is always lower than that for 
the above translational symmetry breaking states with $U'\neq 0$.
Therefore, we focus on the competition between 
CDW and CSAF states in the paper.

\section{Phase transition between the CDW state and the CSAF state}
\label{3}
Let us discuss the three-component fermionic system at finite temperatures.
It has already been clarified that, at zero temperature, 
the CDW and CSAF states are realized in the cases $U<U'$ and $U>U'$,
and are degenerate on the SU(3) symmetric line $U=U'$~\cite{Miyatake}.
Here, we discuss how these phases compete with each other 
at finite temperatures.
Combining DMFT with the NCA method, we calculate the staggered parameters
in the strong coupling region.
\begin{figure}[htb]
  \centering
  \includegraphics[width=7cm]{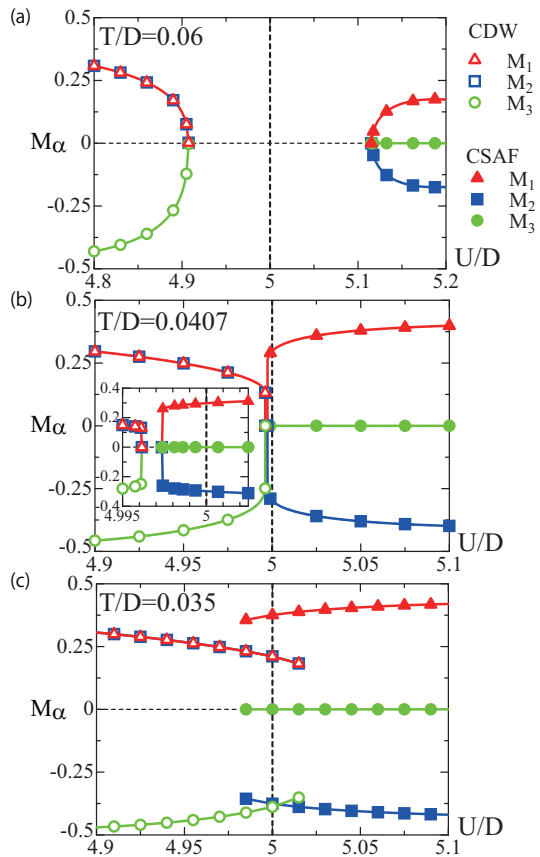}
\caption{(Color online) The order parameters 
as a function of the interaction $U$ 
under the condition $U^\prime/D=-U/D+10$ 
when $T/D=0.06$ (a), $T/D=0.0407$ (b), and $T/D=0.035$ (c). }
\label{OP3}
\end{figure}
Figure~\ref{OP3} shows the results
under the condition $U^\prime/D=-U/D+10$
at finite temperatures.
When $U < U'\; (U/D < 5)$, 
the repulsive interaction between fermions with colors 1 and 2 is 
relatively smaller 
than the others, which favors the CDW state.
In fact, the CDW state is realized with 
$M_1=M_2=0.31$ and $M_3=-0.43$ when $U/D=4.8$
at the high temperature $T/D=0.06$, as shown in Fig.~\ref{OP3}(a).
Increasing $U$ under the condition, 
the magnitudes of the order parameters $|M_\alpha|$ decrease monotonically.
At last, both parameters simultaneously vanish and 
the second order phase transition occurs to the paramagnetic state 
at $(U/D)_c=4.91$, as shown in Fig. \ref{OP3}(a).
On the other hand, the repulsive interaction $U$ tends 
to stabilize the CSAF state.
We find that in the large $U$ region, the CSAF state is indeed realized 
with $M_1=-M_2\neq 0$ and $M_3=0$.
The phase boundary between the CSAF and paramagnetic states 
is obtained as $(U/D)_c=5.12$.
When $U/D=U^\prime/D(=5)$, 
the CDW and CSAF states are degenerate at zero temperature~\cite{Inaba}.
At finite temperatures, magnetic correlations are somehow 
suppressed 
due to a sort of frustration and the paramagnetic state is realized, 
as shown in Fig. \ref{OP3}(a).

Decreasing temperatures, both ordered states become more stable and 
the corresponding phase boundaries approach each other.
At $T/D=0.0407$,
we find in Fig. \ref{OP3}(b) that the CDW and CSAF regions almost 
touch each other around the symmetric point.
Further decreasing temperature, two converged solutions 
appear around the symmetric point $U=U'$ 
in Fig. \ref{OP3}(c).
This implies the existence of the first-order phase transition,
which is consistent with the results at zero temperature~\cite{Inaba}.

\begin{figure}[htb]
\centering
\includegraphics[width=7cm]{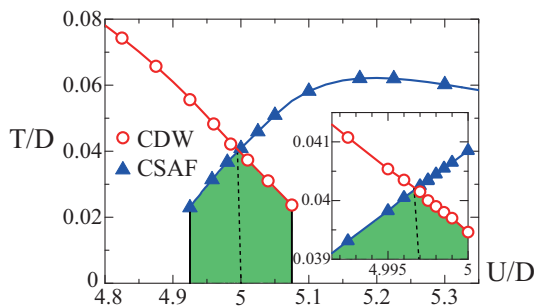}
\caption{(Color online) $U-T$ phase diagram with fixed $U^\prime/D=-U/D+10$.
Circles and triangles represent the phase boundaries 
when the CDW and CSAF solutions vanish.
Two converged solutions exist in the shaded area.
Dashed lines represent the first-order phase boundaries, 
which are guides to eyes.
}
\label{OPT-U}
\end{figure}

The finite temperature phase diagram with fixed $U^\prime/D=-U/D+10$
is shown in Fig. \ref{OPT-U}.
Here, we did not deduce the phase boundary below $T/D=0.02$ since
our method becomes less appropriate in describing low temperature properties
below the critical temperature.
It is found that the CDW (CSAF) state is stabilized in the region with 
$U<U'$ $(U>U')$ at low temperatures.
We also find that both phase boundaries cross each other
at $U/D=4.997$ and $T/D=0.402$.
Below the temperature, the system should have two distinct solutions,
in the shaded region shown in Fig. \ref{OPT-U}.
An important point is that
the crossing point is slightly shifted from the phase boundary 
at zero temperature ($U/D=5$), as shown in the inset of Fig. \ref{OPT-U}.
Therefore, 
the first-order phase boundary between the CSAF and CDW states 
is given by the curve between two points, 
shown as the dashed line in Fig. \ref{OPT-U}.
The phase boundary for the CSAF state 
has the "overhang" structure in the $U-T$ phase diagram,
implying the existence of the reentrant structure.

\begin{figure}[htb]
\centering
\includegraphics[width=7cm]{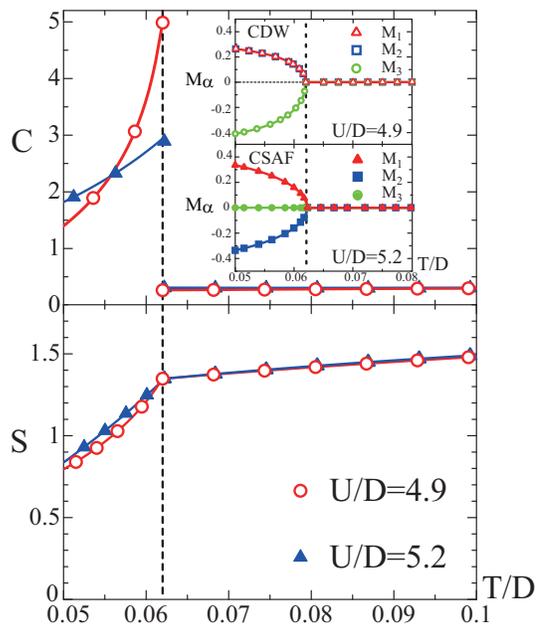}
\caption{(Color online) Specific heat and entropy 
in the system with $U=4.9$ ($5.2$), where its ground state
is the CDW (CSAF) state.
The staggered parameters are shown in the inset.
}
\label{C-S-T}
\end{figure}

The reentrant behavior should originate from the nature of competing states.
To make this point clear,
we calculate the specific heat and entropy for both states,
as shown in Fig \ref{C-S-T}. 
When $U/D=4.9 (5.2)$,
the phase transition occurs to the CDW (CSAF) state
at the critical temperature $T_c/D=0.062$. 
When $T>T_c$, the specific heat and entropy for each case
are almost identical,
as shown in Fig. \ref{C-S-T}.
Decreasing temperatures,
the jump singularity appears in the specific heat
and the cusp singularity appears in the entropy at $T=T_c$.
Note that the singularity in the case $U/D=4.9$ 
is stronger than the other.
This may imply that the entropy in the former case is rapidly released 
just below the transition temperature to stabilize the CDW state, 
while the weaker singularity appears in the latter 
due to the existence of itinerant fermions.
Therefore, if one focuses on the system with $U=U'$, the CSAF state has
the larger entropy at finite temperatures, 
yielding the shift of the first-order phase boundary.

\begin{figure}[htb]
  \centering
  \includegraphics[width=7.5cm]{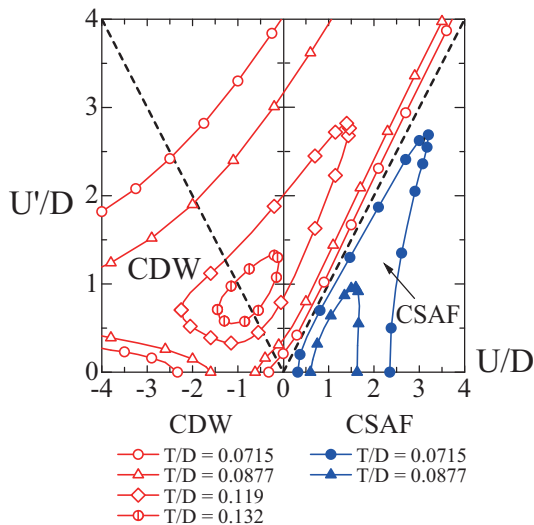}
\caption{(Color online)\, The contour plot of the phase diagram.
Solid (Open) symbols represent the phase boundaries between the CSAF (CDW)
and paramagnetic states.
Dashed lines are the symmetric ones with $U=U'$ and $U=-U'$.}
\label{U-sozu1}
\end{figure}

By performing similar calculations in the $U-U'$ plane ($U'>0$)
at several temperatures, 
we obtain the phase diagram, as shown in Fig. \ref{U-sozu1}.
At zero temperature, the CDW state is realized with $U'>U$ and
the CSAF state is realized with $U'<U$~\cite{Miyatake}.
Increasing temperatures, staggered correlations are suppressed and
the paramagnetic state appears in the weak and strong coupling region,
where the energy scale characteristic of staggered correlations is small. 
In addition, the paramagnetic state appears around symmetric line $U=U'$,
discussed above.
Increasing temperature, 
the CSAF region shrinks toward a certain point on the $U$ axis. 
Since fermions with colors 3 are noninteracting in the case ($U'=0$), 
we can say that the CSAF state is adiabatically connected to 
the two-component antiferromagnetic (AF) state.

On the other hand, the CDW region is widely realized in the phase diagram,
as shown in Fig. \ref{U-sozu1}.
This means that the CDW state is more stable against thermal fluctuations.
Increasing temperatures, the CDW region shrinks toward a certain point 
on $U^\prime=-U$ line.
\begin{figure}[htb]
  \centering
  \includegraphics[width=7cm]{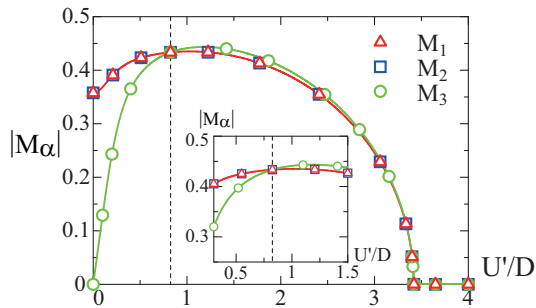}
\caption{(Color online)
The absolute values of the order parameters $|M_\alpha|$ in the CDW state 
when $T/D=0.0715$ and $U/D=-0.825$.
Dashed line indicates the symmetric point $U'/D=-U/D=0.825$.
}
\label{OP-Up}
 \end{figure}
To discuss staggered correlations around the symmetric condition $U'=-U$,
we show in Fig.~\ref{OP-Up} the absolute values of the staggered parameters
in the system with $U/D=-0.825$.
When $U'=0$, the system is decoupled to the interacting fermions 
with colors 1 and 2, and noninteracting fermions with color 3. 
Then, the attractive interaction $U$ stabilizes the density wave state
with $M_1=M_2=0.36$ and no staggered parameter appears in the third component
($M_3=0$).
The introduction of the interaction $U'$ simply stabilizes 
the CDW state, where $|M_3|$ increases rapidly 
and $|M_1|=|M_2|$ also increases.
Therefore, we can say that, in the case, the CDW state is mainly formed by 
the density wave for fermions with colors 1 and 2,
and is regarded as the "$d$-CDW" state.
Increasing the interaction $U'$ beyond the symmetric case $U'=-U$,
we obtain $|M_3| > |M_1|=|M_2|$, as shown in Fig. \ref{OP-Up}.
Since the CDW state is mainly formed by fermions with colors 3,
this state is regarded as the "$s$-CDW" state.
Both magnetizations simultaneously vanish at $U'/D=3.42$,
implying the existence of the second-order phase transition
to the paramagnetic state.
We conclude that, on the symmetric line, 
the crossover between the $s$-CDW and $d$-CDW states
occrurs in the CDW state.


To discuss the maximum critical temperatures for the CSAF and CDW states 
in detail,
we show the finite temperature phase diagram 
with fixed ratios $U'=0$ and $U^\prime=-U$ in Fig.~\ref{3T-U}.
Since the CSAF state in the system with $U'=0$ is equivalent to 
the AF state in the two-component fermion systems,
we have obtained the phase boundary for the two-component Hubbard model.
We find that, in the strong coupling region, the phase boundaries 
for the AF and CDW states are well scaled by {$D^2/4U$ and $3D^2/16U$, 
which are obtained by the second-order perturbation theory.
Therefore, we can say that the NCA solver is appropriate 
in describing the system in the strong coupling region, and 
is expected to yield reasonable results in the crossover region.
In fact, it is found that 
the AF state has a maximum transition temperature 
$T/D= 0.996$ at $U/D=1.65$,
which is consistent with the results obtained from DMFT with CTQMC
method ($T/D=0.1$ and $U/D=2.0$)~\cite{KogaQMC,NCAwerner}.


\begin{figure}[htb]
  \centering
  \includegraphics[width=7cm]{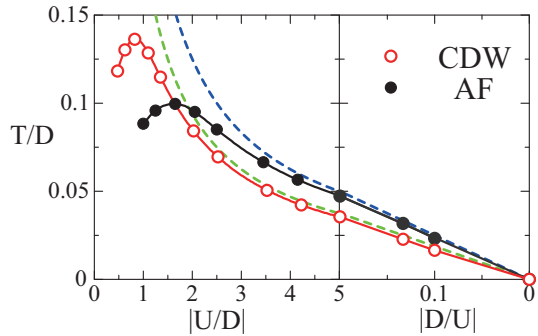}
\caption{(Color online)
Open (closed) circles represent the phase transition temperature for 
the CDW (AF) state with fixed $U^\prime=-U$ $(U'=0)$.
Dashed lines are phase boundaries obtained by means of the second-order
perturbation theory in the strong coupling limit.
}
\label{3T-U}
 \end{figure}
Similar behavior appears in the CDW state, 
where the phase boundary is smoothly changed and 
the crossover occurs between the weak- and strong-coupling states.
The transition temperature for the CDW state 
has a maximum $T/D=0.136$ at $U/D=0.825$.
The above results suggest that the maximum critical temperature for 
the translational 
symmetry breaking state increases with increase in the number of components.
This will be discussed in more detail in the following.

\section{Critical temperatures in the multicomponent fermionic systems}
\label{4}
We here consider multicomponent fermion systems
with the particle-hole symmetric condition
to clarify how the critical temperature depends on the number of components.
This should be important to observe the translational symmetry breaking states
in the fermionic optical lattice.
The model Hamiltonian is given as
\begin{eqnarray}
\hat{\cal{H}}=-t\sum_{\langle i,j \rangle ,\alpha}c^{\dagger}_{i\alpha}c_{j\alpha}+\frac{1}{2}\sum_{\alpha \neq \beta,i}U_{\alpha\beta}n_{i\alpha} n_{i\beta},
\label{eq:multi}
\end{eqnarray}
where $\alpha=1,2,\cdots, N_c$.
Now, we consider the following particle-hole transformations~\cite{Shiba} 
for the $\alpha$th component as
\begin{eqnarray}
c_{i\alpha}&\rightarrow& (-1)^ic_{i\alpha}^\dag,\\
c_{i\beta}&\rightarrow& c_{i\beta},\;\;\;(\beta\neq \alpha).
\end{eqnarray}
Applying it to the model Hamiltonian (\ref{eq:multi}) with $\{U_{\alpha\beta}, U_{\beta\beta'}\}$,
we obtain the model with $\{-U_{\alpha\beta}, U_{\beta\beta'}\}$.
Therefore, these models are equivalent and their ordered states are identified.
For examples, in the ordinary two-component Hubbard model,
the density wave state in the attractive model
is equivalent to the AF state in the repulsive model.
In the three-component system with $U'>0$, 
the CDW state in the model ${\cal H}(t,U,U')$ discussed in the previous section
is equivalent to the trion density wave state in the model ${\cal H}(t,U,-U')$.

We consider the density wave state as one of the
simplest translational symmetry breaking states, 
where $N_c$ fermions are located 
at one of sublattices and the empty site appears in the other.
This state should be realized in the model with the attractive interaction 
$U_{\alpha\beta}=-U(<0)$.
By using the NCA method with $2^{N_c}$ flavors as an impurity solver,
we obtain the maximum temperatures in the system with 
$N_c=2, 3, 4, 5$, and $6$, as shown Fig. \ref{T-U-N}.
\begin{figure}[htb]
\centering \includegraphics[width=7cm]{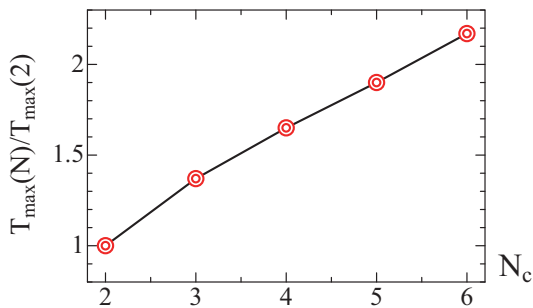}
\caption{(Color online) 
The maximum critical temperature $T_{\rm{max}}$ for the density wave state 
in the $N_c$-component fermionic systems. }
\label{T-U-N}
\end{figure}
We find that $T_{\rm max}$ increases monotonically
with increase in the number of components,
{\it e.g.} $T_{\rm max}(6)/T_{\rm max}(2)=2.2$.

It has been reported that the fermionic optical lattice system with 
two components reaches a low temperature $\sim 1.4T_N$~\cite{1.4T_N}.
Therefore, we expect that the translational symmetry breaking state
should be observed in the fermionic optical lattice system with $N_c > 3$,
{\it e.g.} 
ytterbium atoms with $N_c=6$~\cite{Yb2}, and 
strontium atoms with $N_c=10$~\cite{Sr2}.

The NCA method is one of the powerful solvers to discuss systematically 
finite temperature properties in the system.
However, this method is not useful in describing the system at  
very low temperatures and/or in the weak coupling region quantitatively. 
Nevertheless, the critical temperature for the translational 
symmetry breaking state is relatively higher than
other characteristic temperatures, 
{\it e.g.} Mott and superconducting critical temperatures.
Therefore, we believe that the NCA method is appropriate 
in evaluating the critical temperature
in the crossover and strong coupling regions.

\section{Conclusion}\label{5}
We have investigated the three-component Hubbard model, 
combining DMFT with the NCA method.
We have examined finite temperature properties, 
calculating staggered parameters, specific heat, 
and entropy. 
Then we have found the reentrant behavior in the phase boundary of 
the CSAF state,
which is reflected by the nature of competing ordered states.
We have also discussed the maximum transition temperature 
for two ordered states.
Then we have found that the CDW state is more stable against 
thermal fluctuations and the maximum critical temperature 
is 1.37 times higher than that for the CSAF state.
We have clarified that, increasing the number of components, 
the maximum transition temperature monotonically increases.
For example, the maximum critical temperature in the six-component fermionic 
system is about twice higher than that in the two-component system.
Although the analysis has been performed in the infinite dimensions,
we believe that this tendency is not changed 
even in the three dimensions.
Therefore, the translational symmetry breaking state 
should be observed in the fermionic optical lattice systems with $N_c>3$. 

\begin{acknowledgements}
The authors would like to thank I. Danshita for fruitful discussions.
This work was partly supported by the Grant-in-Aid 
for Scientific Research from JSPS, KAKENHI No. 25800193. (A.K.)
\end{acknowledgements}


\begin{thebibliography}{99}

\bibitem{ultracold1}
I. Bloch, Nat. Phys. {\bf 1}, 23 (2005).

\bibitem{ultracold2}
U. Schneider, L. Hackerm$\rm{\ddot{u}}$ller, S. Will, Th. Best, I. Bloch, 
T. A. Costi, R. W. Helmes, D. Rasch, and A. Rosch,
Science {\bf 322}, (2008) 1520.

\bibitem{ultracold3}
I. Bloch, J. Dalibard, and S. Nascimb$\grave{e}$ne
Nat. Phys. {\bf 8}, 267 (2012).

 
\bibitem{Chin}
J. K. Chin, D. E. Miller, Y. Liu, C. Stan, W. Setiawan, C. Sanner,
K. Xu, and W. Ketterle, Nature (London) {\bf 443}, 961 (2006).

\bibitem{crossover1}
C. A. Regal, M. Greiner, and D. S. Jin
Phys. Rev. Lett. {\bf 92}, 040403 (2004).

\bibitem{crossover2}
M. Bartenstein, A. Altmeyer, S. Riedl, S. Jochim, C. Chin, J. Hecker Denschlag, and R. Grimm
Phys. Rev. Lett. {\bf 92}, 120401 (2004).

\bibitem{crossover4}
J. Kinast, S. L. Hemmer, M. E. Gehm, A. Turlapov, and J. E. Thomas, 
Phys. Rev. Lett. {\bf 92}, 150402 (2004). 

\bibitem{crossover3}
M. W. Zwierlein, C. A. Stan, C. H. Schunck, S. M. F. Raupach, A. J. Kerman, and W. Ketterle
Phys. Rev. Lett. {\bf 92}, 120403 (2004).

\bibitem{crossover5}
M. W. Zwierlein, J. R. Abo-Shaeer , A. Schirotzek , C. H. Schunck, and 
W. Ketterle, Nature {\bf 435}, 1047 (2005).


\bibitem{1.4T_N}
R. A. Hart, P. M. Duarte, T.-L. Yang, X. Liu, T. P. E. Khatami, 
R. T. Scalettar, N. Trivedi, D. A. Huse, R. G. Hulet,
Nature {\bf 519}, 211 (2015).


\bibitem{Li1}
T. B. Ottenstein, T. Lompe, M. Kohnen, A. N. Wenz, and S. Jochim, 
Phys. Rev. Lett. {\bf 101}, 203202 (2008).

\bibitem{Li2}
J. H. Huckans, J. R. Williams, E. L. Hazlett, R. W. Stites, and K. M. O'Hara, 
Phys. Rev. Lett. {\bf 102}, 165302 (2009).

\bibitem{Yb1}
T. Fukuhara, Y. Takasu, M. Kumakura, and Y. Takahashi, 
Phys. Rev. Lett. {\bf 98}, 030401 (2007).

\bibitem{Yb2}
S. Taie, Y. Takasu, S. Sugawa, R. Yamazaki, T. Tsujimoto, 
R. Murakami, and Y. Takahashi, 
Phys. Rev. Lett. {\bf 105}, 190401 (2010).

\bibitem{Sr1}
B. J. DeSalvo, M. Yan, P. G. Mickelson, Y. N. Martinez de Escobar, 
and T. C. Killian, 
Phys. Rev. Lett. {\bf 105}, 030402 (2010).

\bibitem{Sr2}
X. Zhang, M.Bishof, S. L. Bromley, C. V. Kraus, M. S. Safronova, 
P.Zoller, A. M. Rey, J. Ye, Science {\bf 345}, 1467 (2014).



\bibitem{Miyatake}
S. Miyatake, K. Inaba, and S. Suga, 
Phys. Rev. A {\bf 81}, 021603 (2010).


\bibitem{InabaReview}
K. Inaba and S. Suga,
Mod. Phys. Lett. B {\bf 27}, 1330008 (2013).




\bibitem{DMFT1}
W. Metzner and D. Vollhardt, Phys. Rev. Lett. {\bf 62}, 324 (1989).

\bibitem{DMFT2}
A. Georges, G. Kotliar, W. Krauth, and M. J. Rozenberg, 
Rev. Mod. Phys. {\bf 68}, 13 (1996).

\bibitem{DMFT3}
T. Pruschke, M. Jarrell, and J. K. Freericks,
Adv. Phys. {\bf 44}, 187 (1995).


\bibitem{Grewe}
N. Grewe and H. Keiter, Phys. Rev. B {\bf 24}, 4420 (1981).

\bibitem{Kuramoto}
Y. Kuramoto, Z. Phys. B {\bf 53}, 37 (1983).

\bibitem{eckstein}
M. Eckstein and P. Werner, Phys. Rev. B {\bf 82}, 115115 (2010).

\bibitem{Shiba}
H. Shiba, Prog. Theor. Phys. {\bf 48}, 2171 (1972).
\bibitem{InabaMott}
K. Inaba, S. Miyatake, and S. Suga, Phys. Rev. A {\bf 82}, 051602 (2010).


\bibitem{Inaba}
K. Inaba and S. I. Suga, Phys. Rev. Lett. {\bf 108}, 255301 (2012).

\bibitem{Okanami}
Y. Okanami, N. Takemori, and A. Koga, Phys. Rev. A {\bf 89}, 053622 (2014).

\bibitem{Sotnikov1}
A. Sotnikov and W. Hofstetter, Phys. Rev. A {\bf 89}, 063601 (2014).

\bibitem{NCAwerner}
P. Werner, N. Tsuji, and M. Eckstein, Phys. Rev. B {\bf 86}, 205101 (2012).






\bibitem{Twosite}
M. Potthoff, Phys. Rev. B {\bf 64}, 165114 (2001).

\bibitem{SFA}
M. Potthoff, Eur. Phys. J. B {\bf 32}, 429 (2003); {\bf 36}, 335 (2003).


\bibitem{Chitra}
R. Chitra and G. Kotliar, Phys. Rev. Lett. {\bf 83}, 2386 (1999).

\bibitem{KogaQMC}
A. Koga and P. Werner, Phys. Rev. A {\bf 84}, 023638 (2011).





\end{thebibliography}
\end{document}